\documentclass[prb,superscriptaddress,floatfix,twocolumn,showpacs,amsfonts,amsmath,amssymb]{revtex4}
\usepackage{graphicx} 
\usepackage{dcolumn} 
\usepackage{bm} 
\usepackage{color}

\usepackage{tikz}

\begin{document}

\title{A quantum point contact as a (near) perfect spin polariser}

\author{Samuel Bladwell}
\affiliation{School of Physics, University of New South Wales, Sydney 2052, Australia}

\begin{abstract}

In this paper, I present a simple method of obtaining spin-polarised current from a QPC 
with a large Rashba interaction. The origin of this 
spin polarisation 
is the adiabatic evolution of spin ``up" of the first QPC sub-band, into spin ``down" of the 
second QPC sub-band. Unique experimental signatures of this effect can be obtained 
using a magnetic focusing setup, with the characteristic ``double" peak of spin-split magnetic 
focusing only present for conductances $G>2e^2/h$. These particular magnetic focusing 
features are present in recently publish experimental results. Finally, I consider hole QPCs, where 
due to the particular kinematic structure of the Rashba interaction, the required parameters
are much less favourable for experimental realisation. 

\end{abstract}

\maketitle

The prototypical spintronic device is the spin field effect transistor (spin-FET), first proposed by 
Datta and Das in 1990\cite{Datta1990}. The device consists of a spin polarised injector, a gate controllable 
spin-orbit interaction, and a spin sensitive detector. There exist many variations on the spin-FET, but all 
maintain these same essential components.
In Datta and Das original spin-FET proposal, 
ferromagnetic leads were suggested to achieve spin-polarised injection.
More recently, experimental and theoretic 
effort has been devoted to an ``all electric" spin polariser. To this end, elaborate gating potentials 
and many-body effects in quantum point contacts (QPCs) have been studied extensively, with the 
goal of opening a gap between the spin states at the pinch point of the QPC\cite{Debray2009, Kohda2012}. 
The critical experimental 
signature of polarisation being a
``half step" in the conductance, where only one spin state can pass through the constriction
of the QPC.

\begin{figure}[t!]
\begin{center}
     {\includegraphics[width=0.45\textwidth]{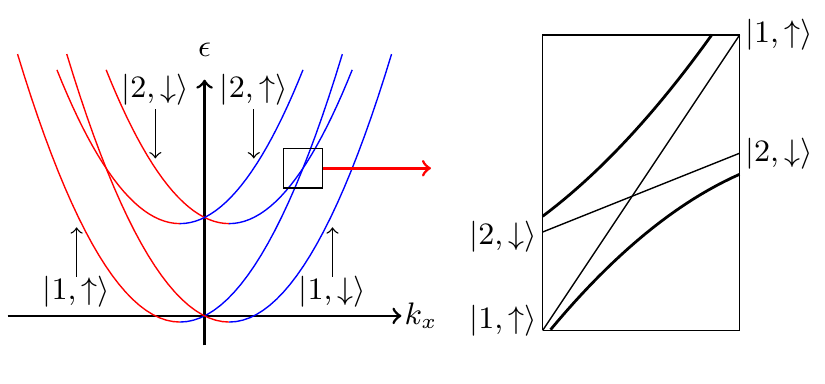}}
	\caption{Right panel: The dispersion of the one dimensional states for the lowest ($n=1$) and 
	second lowest ($n=2$) subbands of the QPC, according to Eq. \eqref{dispersion}. 
	Red (blue) indicates forward (backward) propagating states. At 
	a particular $k_x$, $\left| 1, \uparrow\right>$ and $\left|2, \downarrow \right>$ cross, highlighted in the 
	boxed section. Only the forward propagating states (in blue) are relevant, due to current 
	being injected in one direction. \\
	Left panel: The avoided 
	crossing results in the smooth evolution of $\left| 1, \uparrow\right>$ into $\left|2, \downarrow \right>$.}
\label{cartoon}
\end{center}
\end{figure}

This paper examines an alternative single particle mechanism for generating spin polarised current from QPCs,
without magnetic fields. The spin-polarisation emerges due to an anti-crossing 
between spin ``up" and
``down" states of sub-bands of differing parity, as shown in the left panel of Fig. \ref{cartoon}. 
If the passage through the level crossing is perfectly adiabatic, all ``up" states are converted to 
``down" states, and $100\%$ spin polarised current is obtained. 
This effect
was first proposed and studied numerically by {Eto {\it et al}} with a 
QPC formed in a quantum wire\cite{Eto2005, Eto2006}. 
By varying the length of the QPC 
 constriction and the strength 
of the spin-orbit interaction spin-polarisation of up to 50\% was found.
Similar results have been obtained
elsewhere considering analogous geometries\cite{Reynoso2007, Silvestrov2006}. In Sec. \ref{QPCs}, I present a simple analytical 
model for this effect. In QPCs formed between two reservoirs, the shape of the QPC potential, 
combined with a large, but experimentally achievable, spin-orbit interaction yields (near) perfect 
polarisation. 

The principal difficulty
 of experimental verification of this mechanism of 
spin polarisation is the absence of any features in conductance.
One alternative to conductance measurements is to utilise transverse magnetic focusing,
where strong spin-splitting in momentum results in a real space splitting and 
a ``doubling" of the first focusing peak. The relative height of the spin-split peaks 
can be directly associated with polarisation of the constituent QPCs\cite{Rokhinson2004, Reynoso2007}. 
I show, in Sec. \ref{magfoc}, that
when paired transverse magnetic focusing, 
this particular form of spin polarisation yields a clear experimental signature. This can 
be distinguished from other sources of polarisation by via the QPC conductance dependence.
These unique experimental signatures are present in recently
published results for TMF in InGaAs two dimensional electron systems\cite{Chuang2015, Lo2017}.

Finally,
in Sec. \ref{discuss}, I discuss the case of hole systems. While the spin-orbit interaction in heavy hole 
systems can exceed 30$\%$ of the Fermi energy, the dominant kinematic structure is cubic in momentum\cite{Marcellina2017}. 
This leads to a more complicated form for the spin-orbit interaction in the QPC channel, 
and spin-orbit parameter requirements that are less favourable.

\section{Electron QPC spin polarisation}
\label{QPCs}

\begin{figure}[t!]
\begin{center}
     {\includegraphics[width=0.20\textwidth]{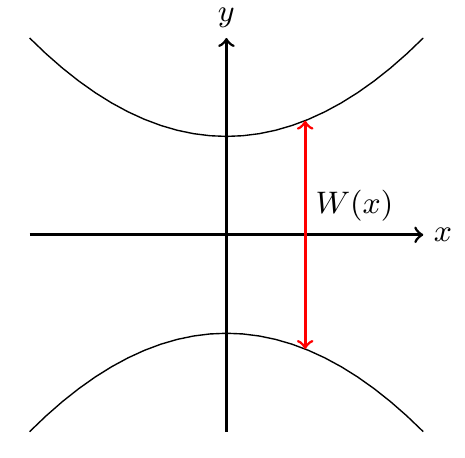}}
	\caption{A typical QPC shape, with varying width. The minimum width, located at $x=0$, is $W_0 = \lambda_F/2$.}
\label{waveguide}
\end{center}
\end{figure}

 A quantum point contact consists of a narrow quasi one-dimensional constriction between two reservoirs. 
 The conductance 
of a QPC occurs in steps of $2e^2/h$\cite{VanWees1988}, and is defined by the adiabatic motion of the one dimensional 
states through the QPC constriction, a result of the smooth variation of the width of the channel\cite{Glazman1988}. 
For simplicity, I consider an infinite square well as the confining potential,
\begin{eqnarray}
{\cal H} = \frac{p_x^2}{2m} + \frac{p_y^2}{2m} + V(y)\label{Ham1} \\ \nonumber
V(y)  = \big\{
\begin{matrix}
0 \quad \text{if } 0<y <W(x) \\
\infty \quad \text{if otherwise}
\end{matrix}
\label{potential}
\end{eqnarray}
with the width $W$ varying adiabatically along the channel; see Fig. \ref{waveguide}. The minimum width of the channel of the QPC is
$W_0 \approx \lambda_F/2$ where $\lambda_F$ is the Fermi wavelength in the reservoirs.
In addition 
to the confinement potential, there is a spin-orbit interaction, 
with interaction strength $\alpha$, 
\begin{eqnarray}
{\cal H}_{so} = \alpha \left( \sigma_x p_y - \sigma_y p_x\right)
\label{Ham2}
\end{eqnarray}
This spin-orbit interaction modifies the one dimensional states, splitting the spin-states in momentum. 
However, there is no change to the transmission coefficient, $T$, and 
hence no change to the conductance curves. 
The dispersion of the one-dimensional states, taking $\left< p_y\right> = 0$, is
\begin{eqnarray}
\varepsilon_n = \frac{\hbar^2 k_x^2}{2m} + \frac{\hbar^2 n^2 \pi^2}{2m W^2} \pm \alpha \hbar k_x
\label{dispersion}
\end{eqnarray}
which is plotted in Fig. \ref{cartoon}. At some characteristic $\alpha$ and $W$ there will be an anti-crossing between the 
spin `up' state of one band, $\left|1, \uparrow \right>$ and the `down state of the band above, $\left|2, \downarrow\right>$. 
In principle, such band crossings occur for arbitrary small $\alpha$, and $\pm k$, however, for current injection, 
only the forward moving states are relevant, the blue components of the dispersion curves in Fig. \ref{cartoon}. 
In a real device, the QPC states project onto the bulk states at 
a finite width, $W_{max}$, and the spin-orbit interaction must be sufficiently large that the width of the crossing is smaller than this. 
Experimental studies place this as being similar to the lithographic width of the device, $W_{max} \sim 200$nm\cite{Molenkamp1990}, 
however, the exact point of decoupling is unknown.  

From Eq. \eqref{dispersion}, this 
crossing point occurs at
\begin{eqnarray}
k_{cross} = \frac{3\hbar \pi^2}{4 m  W^2 \alpha} 
\label{crossing}
\end{eqnarray}
When $k_x > k_{cross}$, the bands are inverted, and  $\left|1, \uparrow \right>$ is above $\left|2, \downarrow\right>$.
The states are well separated in energy far away from the crossing point, and in the crossing 
region, the remaining spin states of the 1st and 2nd sub-band of the QPC are distant. As a result,  the crossing region 
can be described by an 
effective hamiltonian acting on a basis of $\left|1, \uparrow \right>$ and $\left|2, \downarrow\right>$, 
\begin{eqnarray}
{\cal H} = \frac{p_x^2}{2m} + \frac{p_y^2}{2m} + \alpha\sigma_z p_x -  \alpha \sigma_x p_y
\label{effectiveHamiltonian}
\end{eqnarray}
In this crossing region, the linear in $p_y$ term in the Rashba Hamiltonian becomes important, and leads to an 
avoided crossing between $\left|1, \uparrow \right>$ and $\left| 2, \downarrow\right>$, as can be seen in Fig \ref{cartoon}.
The eigenvalues of the effective Hamiltonian, Eq. \eqref{effectiveHamiltonian} are 
 \begin{eqnarray}
&&\varepsilon_\pm =  \frac{\hbar^2 k_x^2}{2m} + \frac{5}{4} \frac{\pi^2 \hbar^2}{m W^2} \pm \Delta \\ \nonumber
 &&\Delta =   \sqrt{\left(\frac{3}{2} \frac{\pi^2 \hbar^2}{2m W^2} - \alpha \hbar k_x \right)^2  +  \left(\frac{8}{3} \frac{\hbar \alpha}{W}\right)^2 }
 \label{splitting}
 \end{eqnarray}
The problem of transitions between $\left|1, \uparrow \right>$ and $\left| 2, \downarrow\right>$ is reduced to 
a standard Landau-Zener problem, with the transition probability, 
\begin{eqnarray}
P = e^{-2\pi \Gamma} \\ \nonumber
\Gamma = \frac{\Delta^2}{2\hbar \frac{d\Delta}{dt}}
\label{LZt}
\end{eqnarray}
where $P$ is the probability of non-adiabatic transitions. 
A critical point to note is the exponential dependence on $\Gamma$ for non-adiabatic transitions,
with a pre-factor of $2 \pi$\cite{Zener1932}. 
Finally, the factor $\Gamma$ can be determined at the 
level crossing itself to be
\begin{eqnarray}
\Gamma  = \frac{4}{3} \frac{m \alpha}{\hbar k_x} \frac{dx}{dW}
\label{adiabatic1}
\end{eqnarray}
In general $k_x < k_F$, where $k_F = \sqrt{2m\varepsilon_F}$ is the 
Fermi momentum in the reservoirs,
\begin{eqnarray}
 {k_x} \approx k_F\left(1 - \frac{5}{3} \frac{m \alpha}{\hbar k_F}\right)
 \end{eqnarray}

$\Gamma$ depends on the shape of the QPC via $dW/dx$. While $dW/dx$ is 
unknown at the crossing point, estimates can be made of its magnitude.
If the width varies 
exponentially, 
\begin{eqnarray}
dW/dx \approx  \frac{W_{crossing}}{L} \ln\left(\frac{W_{max}}{W_0}\right)
\label{grad}
\end{eqnarray}
where $L$ is the length of the QPC. Collimation studies suggest the maximum width
is comparable or slightly smaller than the lithographic 
width of the gates\cite{Molenkamp1990}. The length depends on the lithography of the 
sample, and for a square QPC it is comparable to the exit width. 
The crossing width, $W_{crossing}$ is 
\begin{eqnarray}
W_{crossing} \approx \lambda_F \sqrt{\frac{k_F}{k_\alpha}} \frac{\sqrt{3}}{4} 
\end{eqnarray}
for $m\alpha/\hbar k_F \sim 0.1$, $\lambda_F < W_{crossing} < 3\lambda_F/2$, while the minimum 
width is $W_0 \ge \lambda_F/2$. Taken together, this implies that $dW/dx < 1$ in the crossing region.
 Due to the pre-factor of $2\pi$ in the Landau-Zener transition probability, 
$\Gamma \sim 0.5$ is sufficient to yield near perfect ($\sim90\%$) spin polarisation.

\section{Experimental signatures}
\label{magfoc}

It is important to note that the mechanism of spin-polarisation leaves {\it no}
features in the conductance. This makes detection of spin polarisation from this mechanism a 
difficult exercise. One option is to connect the QPC to ferromagnetic leads; a QPC exhibiting spin-polarisation
will show conductance signatures\cite{Eto2006}. 

Unique experimental signatures can be obtained using transverse magnetic focusing (TMF). A TMF
experiment consists of an injector, and a detector, located in the plane with a weak magnetic field applied 
transverse to the plane of the 2DEG to 
`focus' the electrons from the source to the detector\cite{Tsoi1999}. When the two dimensional charge gas has
a significant spin-splitting, the focusing peaks become spin split, with spin states separated in real space, 
with the splitting,\cite{Bladwell2015}
\begin{eqnarray}
\frac{\delta B}{B} = \frac{2 k_\alpha}{k_F} 
\end{eqnarray}
Spin polarisations can be easily detected as variations in the height of the constituent peaks\cite{Rokhinson2004, Reynoso2007, Chesi2011},  
while complete polarisation will result in the absence of one of the spin-split peaks. 
\footnote{For example, spin-polarisation due to an applied magnetic field will be complete when the injector is tuned below or at $G = e^2/h$, while the double 
peak will be restored when the injector conductance is $G=2e^2/h$. This feature of TMF was demonstrated experimentally
by Rokhinson {\it et al}\cite{Rokhinson2004}. A detailed discussion of this is presented in Ref. \cite{Reynoso2007}}

\begin{figure}[t!]
\begin{center}
     {\includegraphics[width=0.45\textwidth]{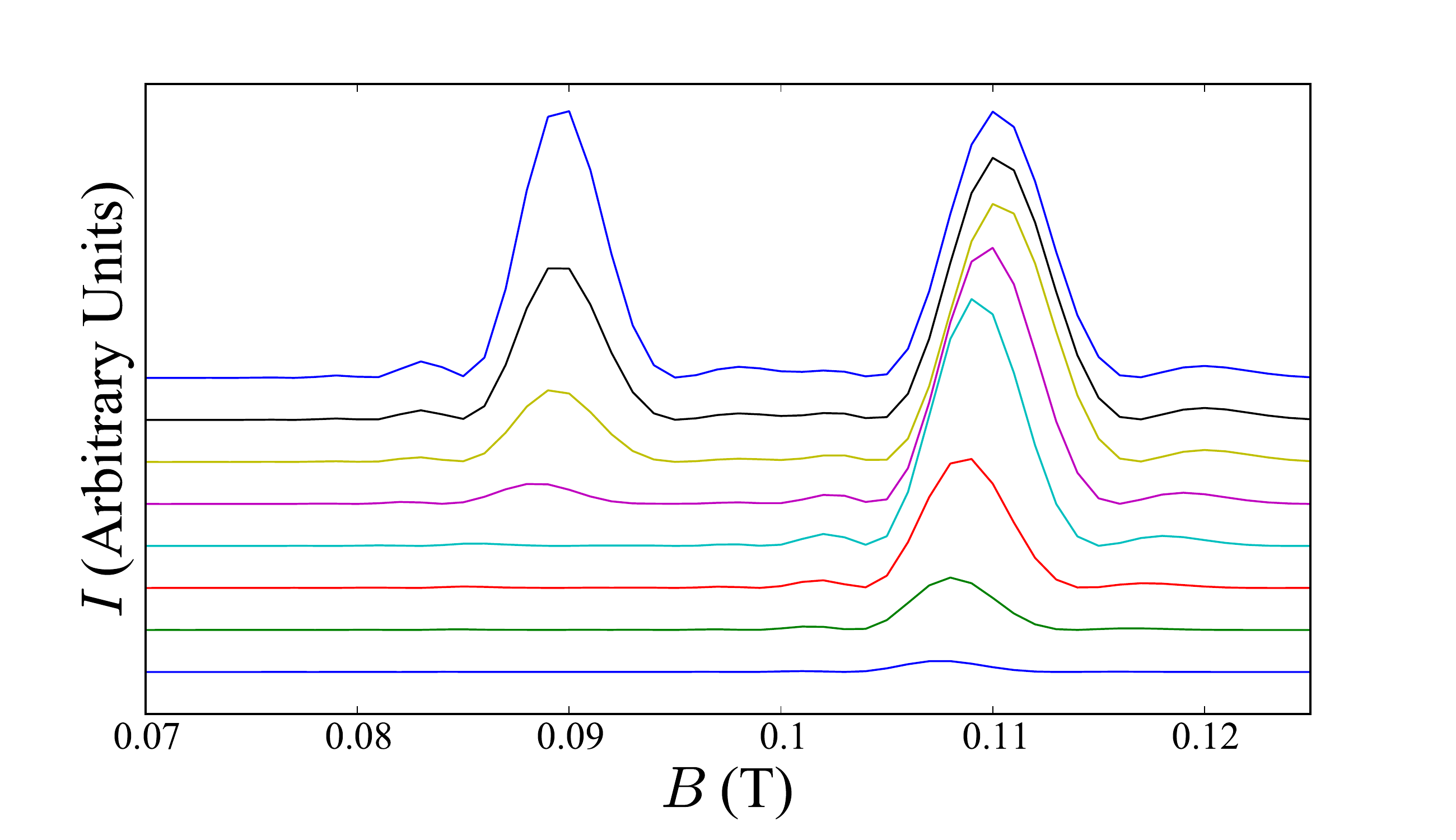}}
	\caption{The magnetic focusing spectrum with an injector QPC with $G = e^2/2h$ to $G=4e^2/h$. The injector 
	is fully polarised. Each curve represents a magnetic focusing trace, the horizontal axis is the magnetic field, 
	the vertical axis is the signal intensity at the detector QPC. 
	The injector QPC conductance increases by $e^2/2h$ for each TMF 
	trace plotted. TMF traces have been offset vertically for clarity. 
	}
\label{theory}
\end{center}
\end{figure}

What makes the TMF signature of adiabatic spin polarisation unique when
compared to polarisation of the QPC due to 
electron-electron interactions or magnetic fields is the continued presence of spin-polarisation when the 
injector is tuned to $G = 2e^2/h$. 
For conductances $G>2e^2/h$, the second TMF peak will gradually appear, with the
peaks only being equal for $G=4e^2/h$, when the lower two bands of the QPC are fully occupied. 
If adiabatic spin-polarisation occurs for $n>1$, the imbalance in the spin-states will persist for arbitrary conductance, with 
finite polarisation for arbitrary conductances. This is unlikely for a QPC with a horn-like shape, since $dW/dx$ 
increases for larger $W$, and the crossing occur at larger values of $W$ for higher bands. 

The adiabatic evolution of the spin states in the injector QPC means that the exit wave-functions contain both the first and 
second mode, 
\begin{eqnarray}
\left| 1, \downarrow\right> \propto \sin\left(\frac{\pi y}{W}\right) \chi_\downarrow \\ \nonumber
\left| 2, \downarrow\right> \propto \sin\left(\frac{2\pi y}{W}\right) \chi_\downarrow
\end{eqnarray}
where $\chi_\downarrow$ is the spin state. While the parity of the states changes, 
the TMF spectrum is insensitive and the shape of the peaks varies only minimally. Using the 
method outlined in Ref. \cite{Bladwell2017}, I have calculated the TMF spectrum, with the
injector exhibiting $100\%$ spin polarisation, and the detector tuned to $4e^2/h$, thereby
allowing both spin-species from either of the sub-bands through. 
For this calculation, I 
use $m\alpha/\hbar k_F =0.125$, with magnetic focusing length of $l = 1500\mu$m, and QPC exit width
of $200$nm. These are approximately inline with the parameters of Ref. \cite{Chuang2015}. 
The plots of the resulting spectrum are presented in Fig. \ref{theory}. 
If the injector is tuned to the second plateau, all sub-bands are occupied, and the double peak structure 
is restored. The TMF spectrum is insensitive to the particular mode structure of the constituent QPCs;
while the angular distribution of the QPCs differs depending on the particular mode\cite{Topinka2000}, the interference spectrum 
is dominated by trajectories close to the phase minimum\cite{Bladwell2017}. 
Additional experimental evidence for this mechanism of spin polarisation could be obtained via this angular dependence, 
which is known to differ significant depending on the particular sub-band of the injecting QPC\cite{Khatua2014, Topinka2001}.

\begin{figure}[t!]
\begin{center}
     {\includegraphics[width=0.45\textwidth]{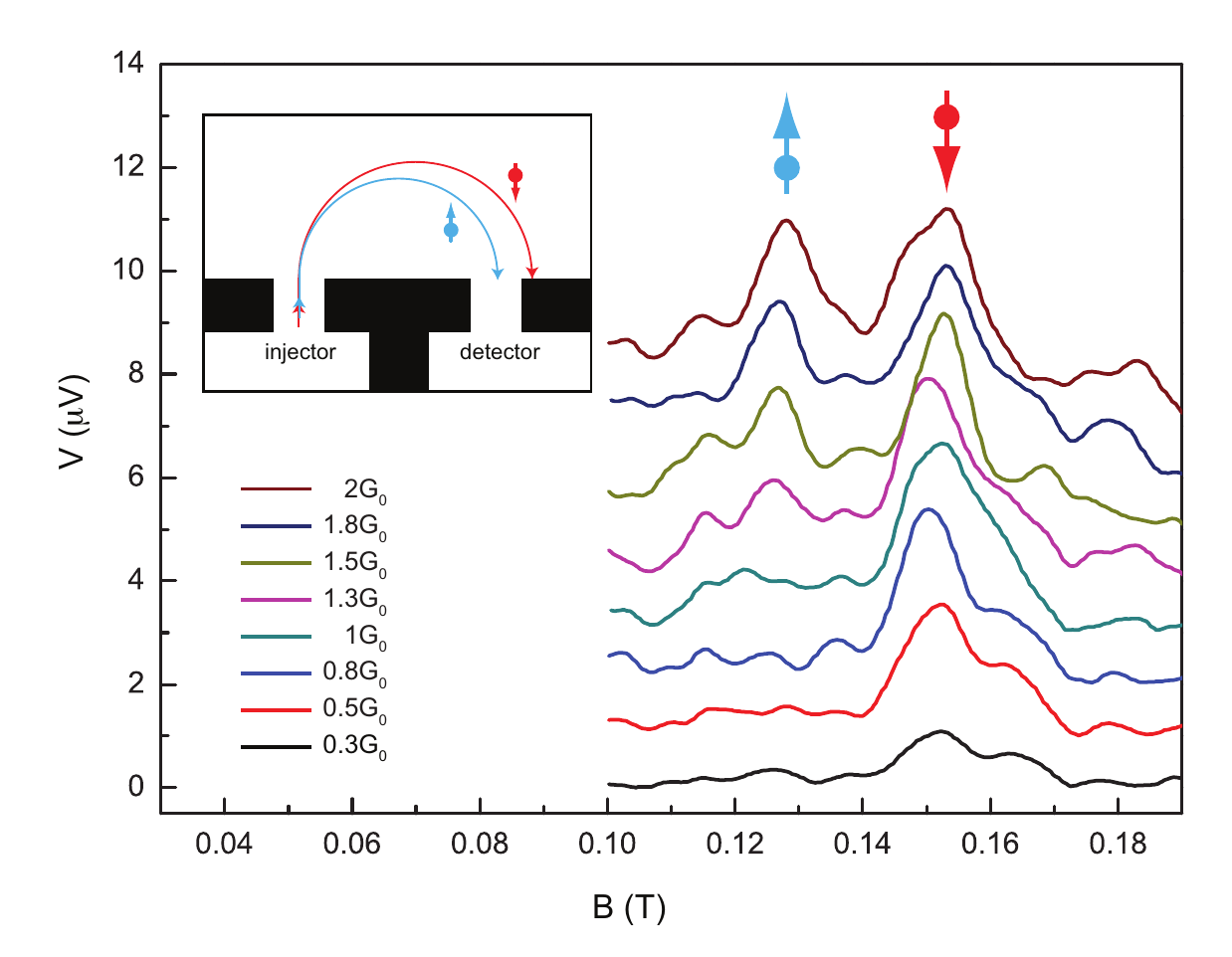}}
	\caption{The magnetic focusing spectrum from Chuang {\it et al}\cite{Chuang2015}. The focusing length is $l\sim 1\mu$ m. A large 
	spin orbit interaction leads to a splitting between the spin-states, with the first magnetic focusing peak doubled. 
	Conductances $G \le G_0$ display only a single peak, a clear signature of adiabatic spin polarisation. The ``doubled" peak 
	structure is only fully restored for $G = 4e^2/h$.  
	}
\label{Chuangresults}
\end{center}
\end{figure}

Recent experimental results from Chuang {\it et al} and Lo {\it et al}, employing
transverse magnetic focusing in a two dimensional electron gas formed in InGaAs\cite{Chuang2015, Lo2017} present
these characteristic features.
The results of Chuang {\it et al} are reproduced in Fig. \ref{Chuangresults}. From the splitting of the focusing 
peaks,
 $m\alpha/\hbar k_F \approx 0.13$. For these results, the side gates of the QPC were biased, resulting 
in an asymmetrical confinement, and an additional spin-orbit interaction $\propto \sigma_z p_x$. 
The effect of this additional spin-orbit interaction is to shift the position of the crossing in the 
QPC, changing $W_{crossing}$. Shifting $W_{crossing}$ can alter $\Gamma$, according to Eqs. \eqref{grad} 
and \eqref{adiabatic1}. 
The detector is tuned to $G = 3e^2/h$, while the injector conductance is varied from $G = 0$ to $G=4e^2/h$. The resulting 
magnetic focusing spectra present spin-polarisation which is persistent to $2e^2/h$, characteristic of 
this mechanism of spin-polarisation. Based on the lack of the double peak structure in the magnetic focusing spectrum 
below $2e^2/h$, the passage through the level crossing is almost completely adiabatic. Other devices
in InGaAs display similar results, with near complete polarisation at $G \approx 3e^2/2h$. The double peak structure 
only restored at $G \approx 3e^2/h$\cite{Lo2017}.

\section{Heavy Hole QPC spin polarisation}
\label{discuss}

\begin{figure}[t!]
\begin{center}
      {\includegraphics[width=0.22\textwidth]{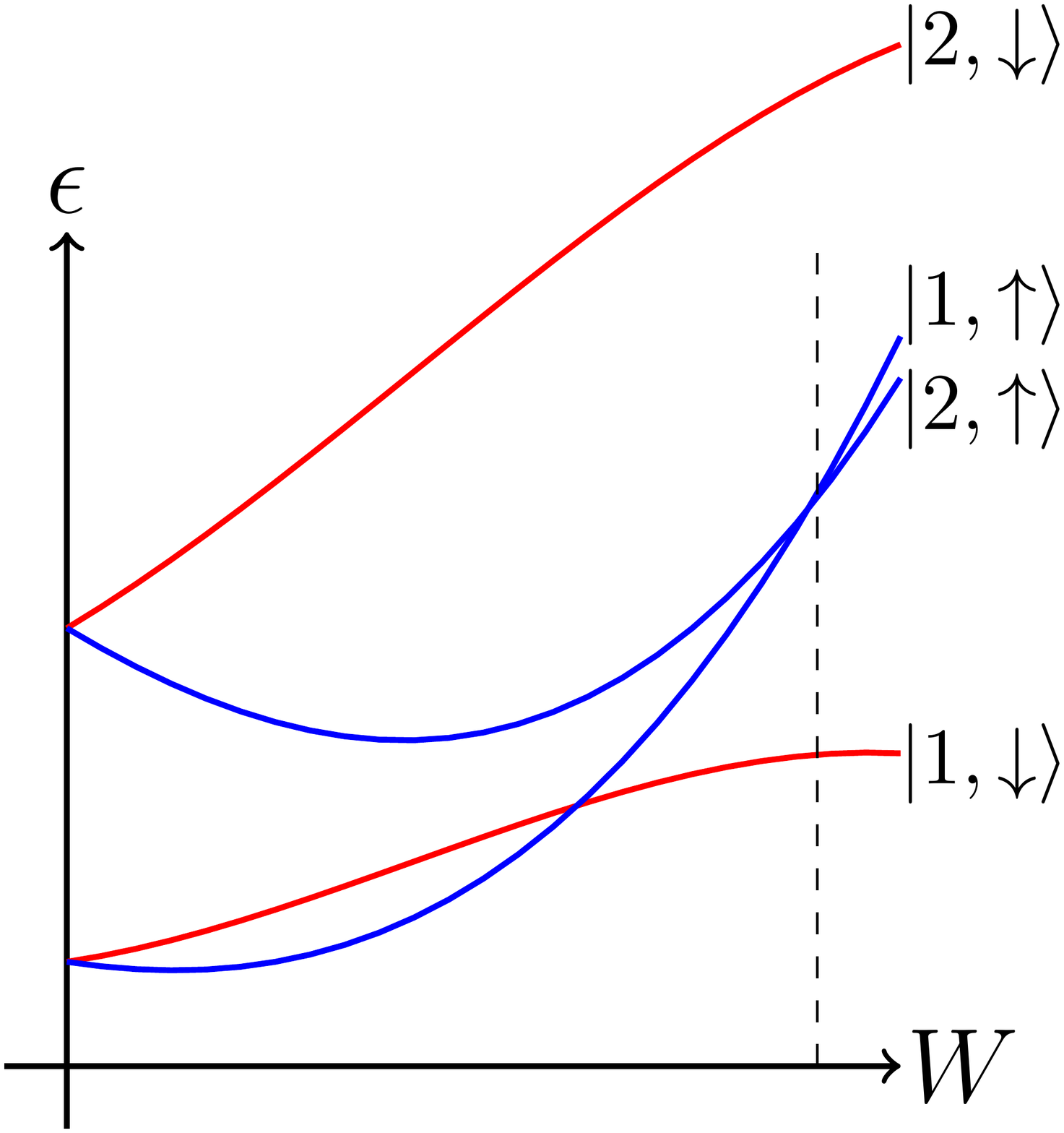}}
     {\includegraphics[width=0.22\textwidth]{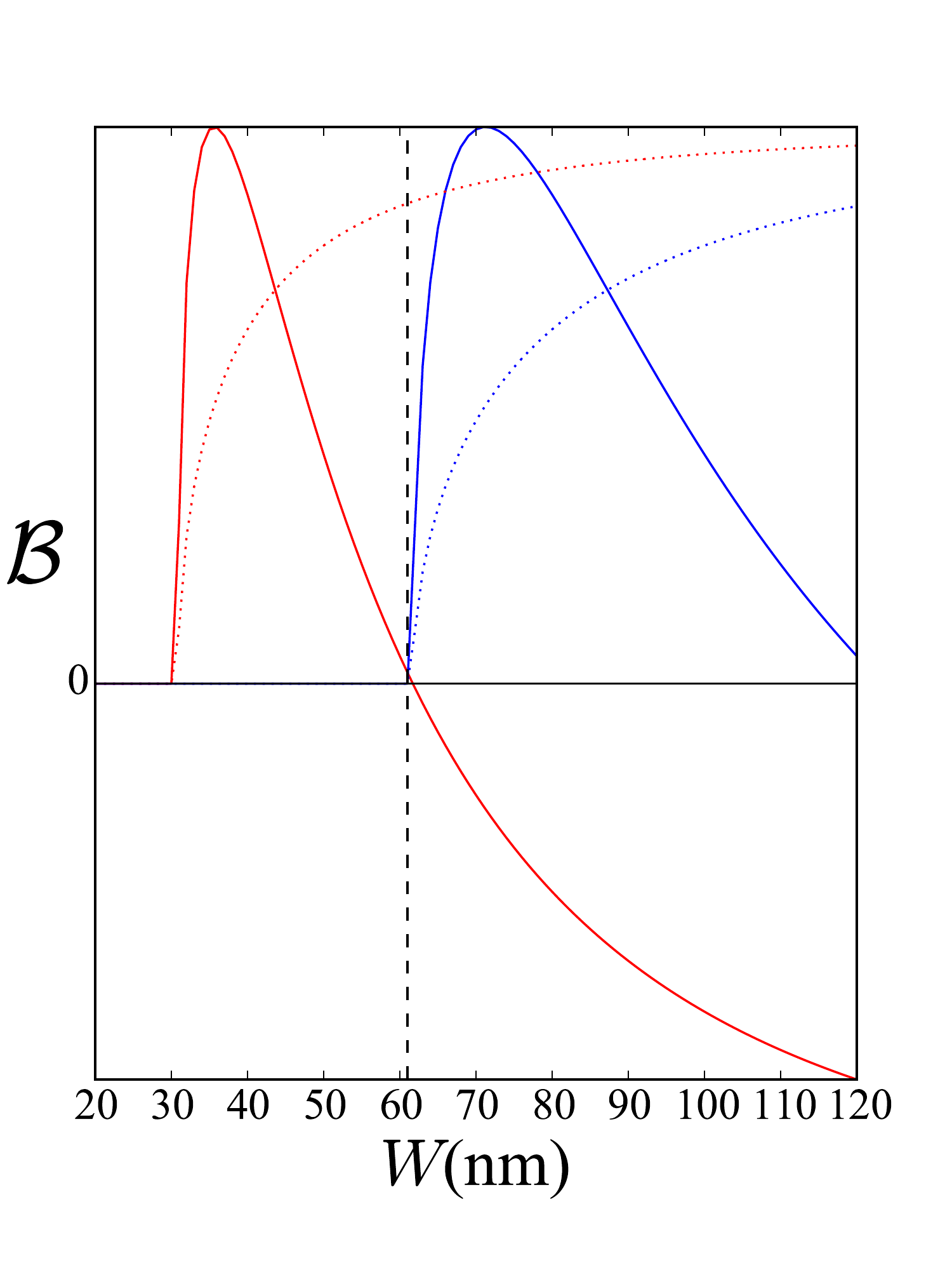}}
	\caption{
	Left Panel: Cartoon of the $n=1$ and $n=2$ sub-bands for heavy holes subject to a Rashba interaction
	in a QPC. Blue and red traces indicate the different spin states. The spin-states are inverted 
	in the $n=2$ sub-band compared to $n=1$. Dashed black line indicates the location of the 
	anti-crossing. \newline
	Right Panel: The strength of the Rashba splitting, Eq. \eqref{cubicrashba}, with varying width 
	at a fixed energy is plotted, with the first sub-band in red, and the second sub-band in blue.
	The vertical dashed line indicates where the second sub-band ($n=2$) is below the Fermi level.
	Critically, at all values where the second sub-band is below the Fermi level, the sign of the 
	interaction opposite the first sub-band.
	For comparison purposes, I have plotted the electron 
 	Rashba strength, from Eq. \eqref{Ham2}, for $n=1$ ($n=2$) in dashed red (blue).
	}
\label{holes}
\end{center}
\end{figure}

Two 
dimensional heavy hole gases can have very large spin-orbit interactions; the Rashba spin-orbit interaction
can exceed $0.3 \varepsilon_F$\cite{Rendell2015}. Due to heavy holes 
having $J_z = \pm3/2$, the dominant  kinematic structure for the Rashba 
interaction is cubic in momentum\cite{Winkler2003, Marcellina2017}, 
\begin{eqnarray}
{\cal H}_{R, h} = \frac{i\gamma}{2} \left(p_+^3 \sigma_- - p_-^3 \sigma_+\right) 
\label{holerashba}
\end{eqnarray}
where $\sigma_\pm = \sigma_x \pm i\sigma_y$, and $p_\pm = p_x \pm i p_y$. Here $\sigma_i$ are 
the Pauli matrices acting on the pseudo-spin doublet $J_z = \pm 3/2$. 
With the confining 
potential of Eq. \eqref{potential}, the kinematic structure of the Rashba interaction is
\begin{eqnarray}
{\cal H}_{R, h} = \gamma \left(3 \left<p_y^2\right> k_x - k_x^3 \right)\sigma_y = {\cal B} \cdot \boldsymbol{\sigma}
\label{cubicrashba}
\end{eqnarray}
I have introduced the effective magnetic field, 
${\cal B}$, which is convenient when discussing the orientation of the spin-states. 
The effective magnetic field, ${\cal B}$,
changes sign at $k_x = \sqrt{3 \left< p_y^2 \right>}$,
shown in Fig. \ref{holes} by the vertically dashed line. 
The change in sign in ${\cal B}$ occurs at different $k_x$ depending on the sub-band, and 
for the $n=1$ sub-band, this coincides with the $n=2$ sub-band first being below the
Fermi energy.
As a result, 
 the effective magnetic field has opposite sign in $n=1$ compared to $n=2$ for a substantial 
 range of widths, $W$. 

Relative to the $n=1$ sub-band, the orientation of the spin-states is ``inverted" in the $n=2$ sub-band. 
That is, $\left|1, \uparrow\right>$ has higher energy than $\left| 1, \downarrow\right>$, 
while $\left|2, \uparrow\right>$ has lower energy than $\left| 2, \downarrow\right>$. While the 
spin-states in the upper band are inverted,
the anti-crossing is between $\left|1, \uparrow\right>$ and $\left|2, \uparrow\right>$, 
as show in the left panel of Fig. \ref{holes}. 
The inversion of the spin-states, and the 
sign change in the effective magnetic field also results in a flipping of the spin filtered 
states in TMF in heavy hole gases\cite{Rokhinson2004, Chesi2011}. 
  
Anti-crossings between opposite spin states which lead to 
polarisation only occur for 
$k_x > 2\pi\sqrt{3}/W$, or $W > \sqrt{3}{\lambda_F}$, when the effective magnetic 
field in the second sub-band changes sign. 
According to Eq. \eqref{grad}, the increase in the crossing width results in a larger $dW/dx$.
 Since $k_x \sim 1/W$, $k_x$ increases at large widths.
Together these increase the Landau-Zener velocity, and therefore suppresses $\Gamma$ in 
Eq. \eqref{LZt}.
As a result, spin-polarisation due to the 
evolution of $\left|1, \uparrow\right>$ into 
$\left| 2, \downarrow\right>$ is considerably weaker in heavy hole gases subject to a Rashba spin-orbit interaction, 
compared to electron systems. 
To compensate for the unfavourable quasi-one dimensional spin-orbit interaction 
in Rashba heavy holes systems smoother QPC geometries could be used.   

 Other kinematic structures are more favourable. The heavy hole Dresselhaus interaction does not exhibit the same 
sub-band dependent sign change as Rashba interaction\cite{Bulaev2005}, 
\begin{eqnarray}
{\cal H}_{D, h} = \beta \left( k_x^2 + \left<p_y^2 \right>\right) k_x \sigma_y
\label{dress}
\end{eqnarray}
where $p_y$ is the transverse momentum. 
The principle limitation in the use of the Dresselhaus interaction 
is the relatively small magnitude; in heterojunctions, $\beta k_F^3 < 0.1 \varepsilon_F$\cite{Marcellina2017}.
Typically Rashba is dominant in heavy hole systems with surface inversion asymmetry, 
and either highly symmetric quantum wells, or other compensating spin-orbit interactions 
would be required.

\section{Summary and Acknowledgements}

In summary, a QPC with a strong spin-orbit interaction can function as a near perfect spin-polarised injector. 
The shape of a typical QPC, combined with a large spin-orbit interaction is sufficient for this to occur. 
Recent transverse magnetic focusing experiments have spectrums with clear evidence
of this particular mechanism of QPC polarisation. While hole systems can have comparable, or even larger
spin-orbit interactions, the kinematic structure of the spin-orbit interaction results in a sign change of the 
spin-orbit interaction for the quasi one dimensional states in the QPC. As a result, the parameters required for 
this mechanism of spin-polarisation in holes are not as experimentally feasible as electrons.

This research was partially supported by the 
Australian Research Council Centre of Excellence in Future Low-Energy Electronics Technologies
 (project number CE170100039) and funded by the Australian Government.
SSRB would like to thank Alex Hamilton, Scott Lilles, and Matt Rendall for valuable comments, and Oleg Sushkov 
for a critical reading and many suggestions for this manuscript.

\end{document}